\documentclass{mn2e}

\def\h0 {$H_0$=70 km s$^{-1}$ Mpc$^{-1}$}

\newcommand{\ce}{\ifmmode {\cal E} \else ${\cal E}$\ \fi}
\newcommand{\kms}{\ifmmode {\rm km\ s}^{-1} \else km s$^{-1}$\ \fi}
\newcommand{\ergs}{\ifmmode {\rm erg\ s}^{-1} \else erg s$^{-1}$\ \fi}

\newcommand{\tes}{\ifmmode \tau_{\rm es} \else $\tau_{\rm es}$\ \fi}
\newcommand{\tk}{\ifmmode \tau_{\rm K} \else $\tau_{\rm K}$\ \fi}
\newcommand{\vfwhm}{\ifmmode V_{\mbox{\tiny FWHM}} \else
            $V_{\mbox{\tiny FWHM}}$\fi}
\newcommand{\msun}{\ifmmode M_{\odot} \else $M_{\odot}$\ \fi}
\newcommand{\afe}{\ifmmode {\mathcal A_{\rm Fe}} \else${\mathcal A_{\rm Fe}}$\ \fi}

\newcommand{\mgii}{Mg {\sc ii}\ }
\newcommand{\civ}{C {\sc iv}\ }
\newcommand{\ciii}{C {\sc iii}\ }
\newcommand{\heiv}{He {\sc iv}\ }
\newcommand{\siiv}{Si {\sc iv}\ }
\newcommand{\ovi}{O {\sc vi}\ }
\newcommand{\nv}{N {\sc v}\ }
\newcommand{\lb}{\ifmmode L_{\rm Bol} \else $L_{\rm Bol}$\ \fi}
\newcommand{\ledd}{\ifmmode L_{\rm Edd} \else $L_{\rm Edd}$\ \fi}
\newcommand{\lx}{\ifmmode L_{\rm 2-10keV} \else  $L_{\rm 2-10keV}$\ \fi}
\newcommand{\hb}{\ifmmode H\beta \else H$\beta$\ \fi}
\newcommand{\ha}{\ifmmode H\alpha \else H$\alpha$\ \fi}
\newcommand{\oiii}{[O {\sc iii}]\ }
\newcommand{\oii}{[O {\sc ii}]\ }

\newcommand{\mbh}{\ifmmode M_{\rm BH}  \else $M_{\rm BH}$\ \fi}
\newcommand{\lv}{\ifmmode \lambda L_{\lambda}(5100\AA) \else $\lambda L_{\lambda}(5100\AA)$\ \fi}
\newcommand{\Msun}{M_{\odot}}
\newcommand{\mdot}{\ifmmode \dot{m} \else \dot{m} \fi }
\newcommand{\llog}{\ifmmode {\rm log} \else {\rm log} \fi }

\usepackage{graphicx}
\usepackage{color}
\usepackage{lscape}

\begin{document}
\title[]{The Origin and Evolution of \civ Baldwin Effect in QSOs from the Sloan Digital Sky Survey}
\author[Y Xu, W. Bian, Q. Yuan, and K. Huang]
{Yan Xu$^{1}$, Wei-Hao Bian$^{1,2}$, Qi-Rong Yuan$^{1}$, and
Ke-Liang Huang$^{1}$ \\
$^{1}$Department of Physics and Institute of Theoretical Physics,
Nanjing Normal University, Nanjing
210097, China\\
$^{2}$Key Laboratory for Particle Astrophysics, Institute of High
Energy Physics, Chinese Academy of Sciences, Beijing 100039,
China\\
}
\maketitle

\begin{abstract}
Using a large sample of 26623 quasars with redshifts in the range
$1.5 \le z\le 5.1$ with \civ $\lambda$1549 \AA\ emission line in
Fifth Data Release of the Sloan Digital Sky Survey (SDSS), we
investigate the cosmological evolution of the Baldwin Effect, i.e.
the relation between the equivalent width (EW) of the \civ
emission line and continuum luminosity. We confirm the earlier
result that there exists a strong correlation between the \civ EW
and the continuum luminosity, and we find that, up to $z\approx
5$, the slope of the Baldwin Effect seems to have no effect of
cosmological evolution. A sub-sample of 13960 quasars with broad
\civ $\lambda$1549 \AA\ emission line from SDSS is used to explore
the origin of the Baldwin Effect. We find that \civ EW have a
strong correlation with the mass of supermassive black hole
(SMBH), and a weak correlation with the Eddington ratio,
$\lb/\ledd$. This suggests that the SMBH mass is probably the
primary drive for the Baldwin Effect.
\end{abstract}


\section{INTRODUCTION}
The discovery of an anti-correlation between the equivalent width
(EW) of \civ $\lambda$1549 emission line and the continuum
luminosity measured at 1450\AA\ in quasar rest frame, known as the
Baldwin Effect, was first made by Baldwin (1977) (e.g. see a review
by Shields 2006). This correlation was explored in Active Galactic
Nuclei (AGNs) for other broad emission lines, such as Ly$\alpha$,
\ciii $ \lambda$1909, \heiv $\lambda$1640, \mgii  $\lambda$ 2800,
\siiv $\lambda$1400, \ovi $\lambda$1034, \nv $\lambda$1240, H$\beta$
$\lambda$4861; for narrow emission lines, such as \oiii$\lambda
5007$, \oii $\lambda 3727$, [N{\sc v}] $\lambda$3426, [N{\sc iii}]
$\lambda$3869; even for the Fe K$\alpha$ emission line in X-ray
spectrum (e.g. Wampler et al. 1984; Baldwin et al. 1989; Netzer et
al. 1992; Borson \& Green 1992; Iwasawa \& Taniguchi 1993; Laor et
al. 1995; Francis et al. 1995; Green 1996; Nandra et al. 1997; Croom
et al. 2002; Dietrich et al. 2002; Baskin \& Laor 2004; Warner et
al. 2004; Page et al. 2004; Zhou et al. 2005; Netzer et al. 2006).
Over the past 30 years, a significant amount of effort has been
expended to confirm this effect and explore its origin and
evolution.

The slope of line EW versus the luminosity for different ions
depends on their ionization energy (e.g. Dietrich et al. 2002).
However we are not sure whether the slope of the Baldwin Effect
evolves with redshift.

Several interpretations about the origin of Baldwin effect have
been proposed. It is not clear whether the Baldwin effect is
driven by the underlying physical parameters, such as the
redshift, Eigenvector 1 of Boroson \& Green (1992), the Eddington
ratio (i.e. the ratio of the bolometric luminosity to the
Eddington luminosity), the mass of supermassive black hole (SMBH),
the gas metallically (e.g. Dietrich et al. 2002; Shang et al.
2003; Sulentic et al. 2007; Baskin \& Laor, 2004; Bachev et al.
2004; Warner et al. 2004). Baskin \& Laor (2004) found a strong
correlation between the \civ EW and the Eddington ratio, and
suggested that the Eddington ratio is the primary physical
parameters driving the Baldwin effect. However, Shang et al.
(2003) used the method of Spectral Principal Component (SPC) to
find there is no correlation between the Baldwin effect and the
Eddington ratio or the mass as underlying physical parameters.

With the reverberation mapping result, we may calculate the virial
SMBH mass from the broad lines (e.g., \hb, \ha, \mgii, \civ) (e.g.
Kaspi et al. 2000; Bian \& Zhao 2004; Vestergaard \& Peterson
2006; Salviander et al. 2007). It provides the possibility to
explore these underlying physical parameters for the origin of
\civ Baldwin effect.

In the past, most of the work was made by using relatively small
samples. For a sample of 20 quasars, Baldwin (1977) first
discovered this effect. Wills et al. (1999) used a complete sample
of 22 QSOs from the Bright Quasar Survey (LBQS) to discuss the
relation between the Baldwin effect and the Eigenvector 1. For a
sample of 993 quasars from the Large Bright Quasar Survey (LBQS),
Green, Forster \& Kuraszkiewicz (2001) studied the relations of
\civ EW with the continuum luminosity and redshift \footnote
{Their luminosity at 2500 \AA is derived from the $B_{\rm J}$
photometric magnitude, since the LBQS spectra are not
spectrophotometric}. Dietrich et al. (2002) used a large sample of
744 type 1 AGNs ($0< z< 5$) from International Ultraviolet
Explorer (IUE) and Hubble Space Telescope (HST) to discuss the
correlations of continuum and emission-line properties in the
rest-frame ultraviolet and optical spectral ranges.

Here we use a large sample of 26623 quasars with \civ emission
line from the Fifth Data Release (DR5, Adelmann et al. 2007)
catalog of the Sloan Digital Sky Survey (SDSS) to discuss \civ
Baldwin Effect and its evolutionary effect, and we also use a
large sample of 13960 quasars with broad \civ emission line to
discuss the underlying physical parameters of the Baldwin Effect.
The sample is described in \S 2, the results and the analysis are
given in \S 3, and the conclusions are presented in \S 4.

\section{Sample}
The sample used in this paper is selected from SDSS-I DR5
(Adelmann et al, 2007). SDSS-I was completed and DR5
(Adelman-McCarthy et al. 2007) was made public on 30 June 2006.
The SDSS Quasar Survey is continuing via the SDSS-II Legacy
Survey. SDSS-I DR5 covers an imaging area of about 8000 square
degrees and a spectroscopic area of about 5740 square degrees.
SDSS DR5 catalog consists of 90,611 objects with automated
spectral classification in SDSS DR5. Recently, Schneider et al.
(2007) presented a quasars catalog from SDSS DR5, consisting of
77,429 quasars with $M_{i}<-22$. The wavelength coverage of SDSS
spectrum is 3800\AA\ to 9200\AA\, and we select 33943 quasars with
redshift $1.5 \le z \le 6.0$ to make \civ $\lambda$ 1549 available
in SDSS spectrum.

There are three stages during the pipeline of SDSS spectral fit. The
first two stages are used to find lines to determine the line
redshift. The third stage is used to measure the line feature by a
single Gaussian fit for the continuum subtracted spectrum. The
chi-squared values can be used to evaluate the quality of the fit.
\footnote {http://cas.sdss.org/astrodr5/en/help/docs/algorithm.asp}
We select the following parameters of these quasars for our
discussion: redshift (\texttt{z}), mjd of observation
(\texttt{mjd}), plate ID (\texttt{plate}), fiber ID
(\texttt{fiberID}), $\sigma$ of fitted gaussian (\texttt{sigma}),
error of $\sigma$ (\texttt{sigma\_Err}), equivalent width (EW),
error in equivalent width (\texttt{EW\_Err}), continuum flux value
at 1549\AA ($f_{1549}$), $\chi^2$ of fit (\texttt{chisq}), and
degree of freedom (\texttt{nu}, $\nu$).

Through the SQL search language \footnote
{http://cas.sdss.org/astro/en/help/docs/sql\_help.asp}, we get 27545
objects from SpecLine table in the SDSS DR5 with automated spectral
classification of quasar, redshift $1.5 \le z \le 6.0$, and the
redshift confidence not less than 95\%.

922 objects with negative values of \civ EW or with
\texttt{EW\_Err} values larger than the values of its EW are
excluded. At last, we have a sample of 26623 quasars with \civ
emission line to discuss the \civ Baldwin Effect and its
evolutionary effect. In Figure 1 we show the histogram of the
redshift. Only two quasars have redshift larger than five, their
reshifts are 5.032.

In order to explore whether the Eddington ratio or/and the SMBH mass
is/are the underlying physical parameters for the origin of the
Baldwin Effect, we calculate the SMBH mass for quasars with broad
\civ line. For a sample of 130 AGNs with HST archival spectra,
Sulentic et al. (2007) suggested subtracting a narrow component of
\civ $\lambda$1549 \AA\ with FWHM $\le 1500 \kms$ to discuss \civ
$\lambda$1549 \AA\ as an Eigenvector 1 parameter. Then in our paper,
from SDSS DR5, a sub-sample of 13960 quasars with broad \civ
emission line is selected by the following criterions:
(1)$\sigma/(1+z)\ge 3.3$\AA\ , corresponding 1500\kms of \civ FWHM,
(2)$\chi^2/\nu < 1.5$, showing acceptable one gaussian fit of \civ
emission line.

\begin{figure}
\begin{center}
\includegraphics[width=8cm]{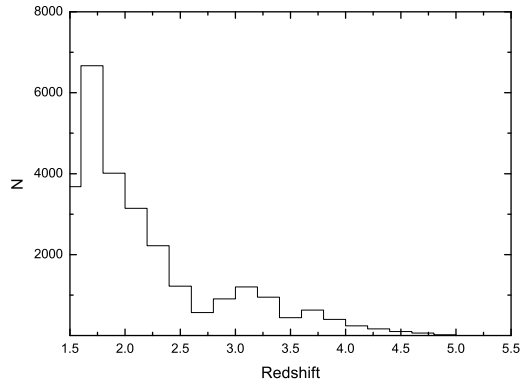}
\caption{Redshift distribution for the 26623 quasars.}
\end{center}
\end{figure}

\section{RESULTS AND DISCUSSION}
\subsection{The slope and correlation
coefficient of the Baldwin Effect}

\begin{figure}
\begin{center}
\includegraphics[width=9cm]{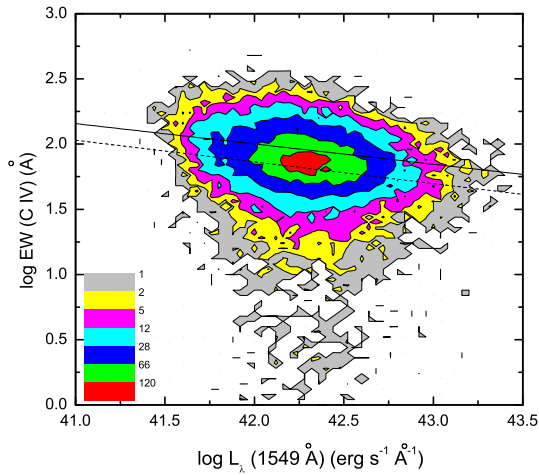}
\caption{\civ EW (in units of \AA) versus the continuum luminosity
at 1549 \AA (in \ergs \AA$^{-1}$) for total 26623 quasars. The
solid line is the least-square linear fit considering the error of
\civ EW. The dash line is the least-square linear fit without
considering the error of \civ EW.}
\end{center}
\end{figure}

The \civ EW is from the result of SQL searching. The continuum
luminosity $L_{\lambda}(1549 \AA)$ at rest frame is calculate from
its flux ($f_{1549}$) by the following formula,

\begin{eqnarray}
L_{\lambda}(1549\AA)= 4\pi d_{\rm L}^2 f_{1549}
\end{eqnarray}
where $d_{\rm L}$ is the luminosity distance. For flat universe,
\begin{eqnarray}
d_{\rm L}=c(1+z)H_0^{-1} {\int^{z}_{0}dz[(1+z)^2(1+\Omega _{\rm
M}z)-z(2+z)\Omega _{\rm \Lambda}]^{-1/2}}\}
\end{eqnarray}
where $c$ is the speed of light, $\rm H_0$ is the present Hubble
constant, $\Omega_{\rm M}$ is the mass density, $\Omega_{\rm
\Lambda}$ is the vacuum density. The cosmological constants we use
in our paper are $\Omega _{\rm \Lambda}=0.7$, $\Omega _{\rm M}=0.3$,
$\rm H_{0}=71 \kms$.

The following formula is used to express the relation between \civ
EW and the continuum luminosity,
\begin{eqnarray}
\log \rm EW{\mbox{(\rm \civ)}}=\alpha+\beta \log L_{\lambda}(1549
\AA)
\end{eqnarray}

Figure 2 gives the correlation between \civ EW and continuum
luminosity at 1549\AA\ for 26623 quasars from SDSS DR5. Taking the
average of the upper and lower errors of \civ EW as the weight, we
use the weighted least-square linear regression to parameterize
the strong correlation (Press et al. 1992, P. 655). The
correlation coefficient is $R=-0.265$ and the null hypothesis is
less then $10^{-4}$. The slope is $\beta=-0.156\pm 0.0002$ and the
intercept is $\alpha=8.549\pm 0.011$. An unweighted least-square
fitting is also given (denoted with dash line in Figure 2). We
notice that the centroid of the distribution of the data is
between these two fit lines. The dash line is shifted vertically
with respect to the solid line, which is due to that the \civ
lines with lower EWs usually have the larger errors in logarithm
space, and thus have smaller weights in our fitting. In the
hereafter linear fittings, we just show the results of weighted
least-square linear fittings, taking the mean value of the upper
and lower errors of \civ EWs as the fitting weights.

By IUE spectra, Kinney et al. (1990) found that the slope is
$-0.17\pm 0.04$ and Espey \& Andreadis (1999) found it is
$-0.17\pm 0.03$. By the optical selected sample, Zamorani et al
(1992) found it is $-0.13\pm 0.03$. For a sample of 158 AGNs
observed with the Faint Object Spectrograph on the Hubble Space
Telescope, Kuraszkiewicz et al. (2002) found it is $-0.198\pm
0.036$. Dietrich et al. (2002) found the slope is $-0.14\pm0.02$
for their total sample, and $-0.20\pm 0.03$ for objects with $\log
\lambda L_{\lambda}(1450 \AA) \ge 44~ \ergs$. We get a slope,
$\beta=-0.156\pm 0.0002$, which is consistent with some previous
studies (Osmer et al. 1994; Laor et al. 1995; and Green 1996;
Dietrich et al. 2002). In table 1, we list the results of the
above researches.

The redshift coverage of our sample is from 1.5 to 5.1 (see Figure
1). We divide our sample into four parts according to the redshift,
and the redshift bins are: $1.5 \le z \le 2$, $2<z \le2.5$, $2.5<
z\le 3.5$, and $3.5< z\le 5.1$.  Figure 3 presents the \civ Baldwin
Effect in different redshift bins. For each redshift bins,
considering the error of \civ EW, we use the least-square linear
regression to find the correlation between \civ EW and the continuum
luminosity at 1549\AA. We find a strong correlation for different
redshift bins. The correlation coefficients and the slopes are
listed in Table 2.  Steeper slopes are found in these four
redshift-bins, respect to the total sample, and this is similar to
the results of the paper from Dietrich et al. (2002). For these four
bins, the slopes are almost the same, which shows no cosmological
evolution of the \civ Baldwin Effect up to $z\approx 5$.

\begin{figure*}
\begin{center}
\includegraphics[width=20cm]{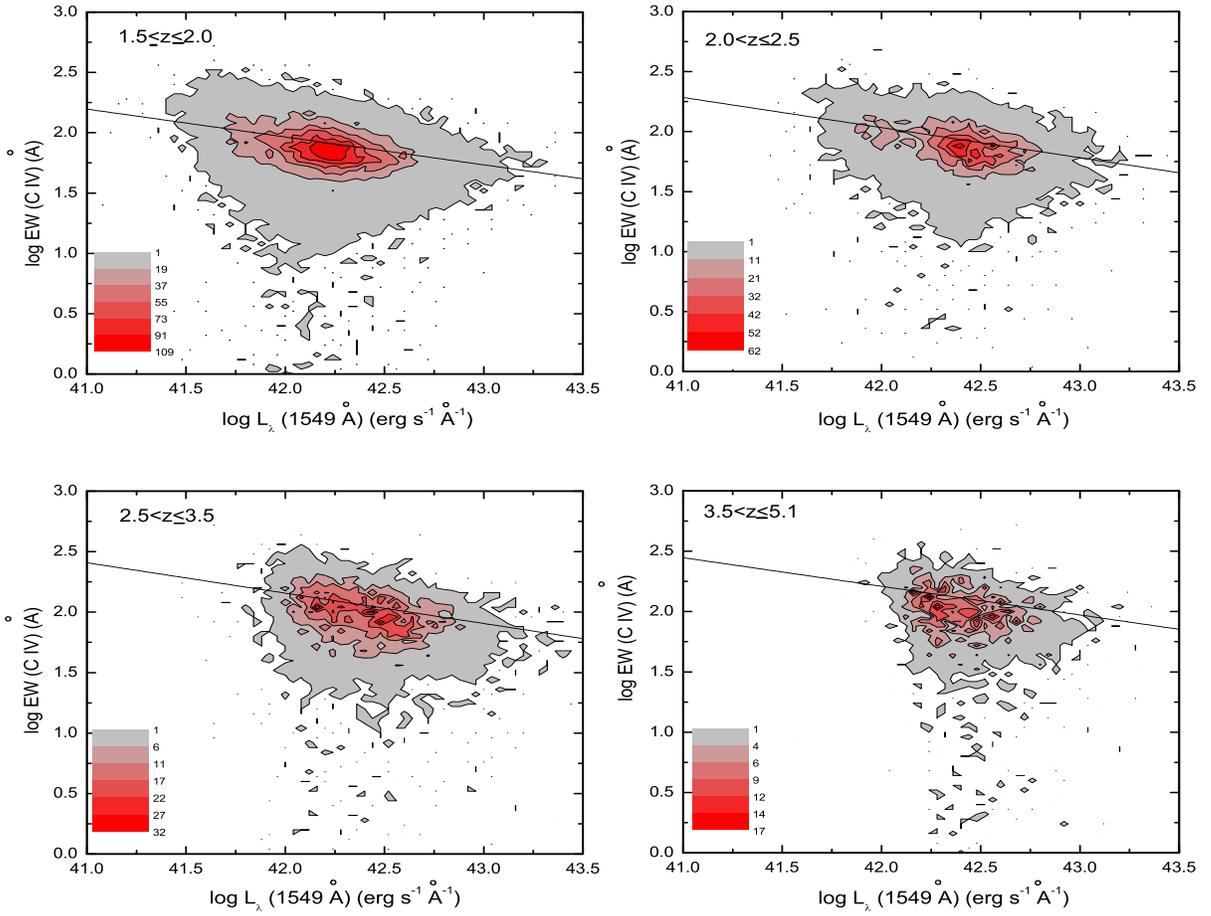}
\caption{log-log correlation between the \civ EW and the continuum
luminosity at 1549\AA\ (in \ergs \AA $^{-1}$) for the DR5 sample
of 26623 quasars in different redshift bins. Each redshift bin is
marked in every graph. Solid lines are the linear fits.}
\end{center}
\end{figure*}

\begin{table*}
\centering \caption{Results of the previous studies and of our
work. Col.(1): the article; Col.(2): the sample of the work;
Col.(3): redshift; Col.(4): the slopes of the linear fit for the
Baldwin Effect of the studies.}
\begin{tabular}{lcccccccccc} \hline \hline
article  & sample  & redshift  & slope  \\
(1) & (2) & (3) & (4)  \\
\hline our paper & 26623 quasars from SDSS DR5 & $1.5 \le z < 5.1$
& $-0.156\pm 0.0002$\\
Kinney et al.(1990) & 101 quasars and 88 Seyfert 1 galaxies & $z <
2.0$ & $-0.17\pm 0.04$\\
Espey et al.(1999) & ~200 AGNs & $z > 1.0$ & $-0.17 \pm 0.03$\\
Zamorani et al.(1992) & five 'complete' samples &  & $-0.13 \pm
0.03$\\
Kuraszkiewicz et al.(2002) & 158 AGNs & $0.001 \le z \le 3.822$ &
$-0.198 \pm 0.036$\\
Dietrich et al.(2002) & 744 AGNs & $0 \le z \le 5.0$ & $-0.14\pm
0.02 $ \\
\hline
\end{tabular}
\end{table*}

\begin{table*}
\centering \caption{Results for the samples in different redshift
bins. Col.(1): different redshift bins; Col.(2): number of quasars
in different redshift bins; Col.(3): correlation coefficient of
the linear fit; Col.(4): slope of the linear fit together with its
error; Col.(5): intercept of the linear fit with its error at $log
L_ \lambda (1549\AA)=10^{42}$ \ergs; Col.(6): the possibility of
null hypothesis. }
\begin{tabular}{lccccccccccc} \hline \hline
z  & N  & R  & slope & intercept  &P \\
(1) & (2) & (3) & (4) & (5) & (6) \\
\hline
$1.5\le z\le 2.0$  & 14370 & $-0.360$ & $ -0.231 \pm 0.0003 $ &$ 1.980 \pm 0.022 $ & $ <10^{-4}$ \\
$2.0 < z \le 2.5$  & 6076  & $-0.349$ & $ -0.251 \pm 0.0005 $ &$ 2.050 \pm 0.032 $ & $ <10^{-4}$ \\
$2.5 < z \le 3.5$  & 4338  & $-0.384$ & $ -0.251 \pm 0.0006 $ &$ 2.176 \pm 0.035 $ & $ <10^{-4}$ \\
$3.5 < z \le 5.1$  & 1839  & $-0.337$ & $ -0.237 \pm 0.0015 $ &$ 2.210 \pm 0.092 $ & $ <10^{-4}$ \\
\hline
\end{tabular}
\end{table*}

\subsection{The origin of Baldwin Effect}
The physical origin for the Baldwin Effect is still an open
question. It is suggested that the Baldwin Effect is driven by the
Eddington ratio and/or the SMBH mass (e.g. Baskin \& Laor 2004;
Warner et al. 2004). Using 81 BQS quasars, Baskin \& Laor (2004)
found that the correlation coefficient between \civ EW and the
continuum luminosity is $-0.154$ , and the correlation coefficient
between \civ EW and the Eddington ratio is $-0.581$. They
suggested that Baldwin Effect is driven by Eddington ratio,
$\lb/\ledd$.

We calculate the SMBH mass from FWHM of \civ emission line by the
following formula (e.g. Vestergaard \& Peterson 2006),
\begin{eqnarray}
\log \,M_{\rm BH} \mbox{(\rm \civ)} =~~~~~~~~~~~~~~~~~~~~~~~~~~~~~~~~  \\
\nonumber
   \log \,\left[ \left(\rm \frac{FWHM\mbox{(\rm \civ)}}{1000~km~s^{-1}} \right)^2 ~
   \left( \frac{\lambda L_{\lambda} (1350\,{\rm \AA})}{10^{44} \rm erg~s^{-
1}}\right)^{0.53}
    \right]
     + (6.66 \pm 0.01).
\end{eqnarray}
where $\rm FWHM(\mbox{\rm \civ})=2.35\sigma/(1+z)$ in units of
\kms, $z$ is the redshift. The uncertainty of the SMBH mass is due
to the errors of $\rm FWHM(\mbox{\rm \civ})$, $\lambda L_{\lambda}
(1350\,{\rm \AA})$, and the system error in equations (4) is from
the uncertainties of the BLRs geometry and dynamics. The
uncertainty of our calculated SMBH mass is about 0.5 dex
(Vestergaard \& Peterson 2006).

We also calculate the Eddington ratio, i.g. the ratio of the
bolometric luminosity (\lb) to the Eddington luminosity (\ledd),
where $\ledd =1.26 \times 10^{38} (M_{\rm BH}/ \msun) \ergs$. The
bolometric luminosity is calculated from the monochromatic
luminosity at 1350\AA\ , $\lb =c \times L_{1350}$ , where
$L_{1350}$ is the continuum luminosity at 1350\AA\  and $c=3\sim
5$ (Ricahrads et al. 2006), and we take $c=4$. Considering the
uncertainty of our calculated SMBH mass is about 0.5 dex, the
uncertainty of the Eddington ratio is about 0.5 dex or more.

In order to calculate the SMBH mass, we use a sub-sample of 13960
quasars with broad \civ emission line from SDSS DR5, which is
stated in section 2. In Figure 4, we show the relations between
the \civ EW with SMBH mass, the continuum luminosity at 1549 \AA\
, and Eddington ratio, $\lb/\ledd$. The results of these three
relations are listed in Table 2.

Our results show that the correlation coefficient for the relation
between \civ EW and the continuum luminosity is larger than the
other two relations. And the correlation coefficient of \civ EW
with SMBH mass is larger than that of \civ EW with Eddington
ratio. Therefore, by the larger sample, we can't confirm the
suggestion that the Eddington ratio is the underlying physical
parameter for \civ Baldwin Effect, and maybe the SMBH mass is a
better driving parameter.

\begin{figure*}
\begin{center}
\includegraphics[width=11cm]{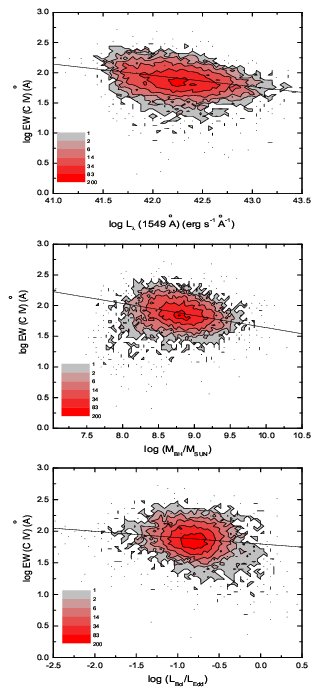}
\caption{the log-log correlation between the \civ EW with SMBH
mass, the Eddington ratio, $\lb/\ledd$, and the continuum
luminosity for the 13960 DR5 sample. The solid line is the linear
fit.}
\end{center}
\end{figure*}

\begin{table*}
\centering \caption{Results for different relations for the 13960
quasars with broad \civ emission line. Col.(1): different
correlations; Col.(2): correlation coefficient of the linear fit
for the different correlations; Col.(3): slope of the linear fit
together with the error; Col.(4): intercepts of the linear fit
with the errors at $\rm{log} L_ \lambda (1549\AA)=10^{42}$ \ergs,
$\rm{log} M_{\rm BH}/\Msun = 9$, $\rm{log} \lb/\ledd=-1$,
respectively; Col.(5): the possibility of null hypothesis.}
\begin{tabular}{lcccccccccc} \hline \hline
relations & R & slope & intercept & P\\
(1) & (2) & (3) & (4) & (5)\\
\hline
log \civ EW - log $ L_ \lambda$ (1549\AA) & $-0.372$ & $-0.187\pm 0.0004$ & $1.954 \pm 0.023$& $<10$ $^{-4}$ \\
log \civ EW - log $ M_{\rm BH}/\Msun$     & $-0.344$ & $-0.198\pm 0.0005$ & $1.853 \pm 0.006$& $<10$ $^{-4}$ \\
log \civ EW - log $\lb/\ledd$             & $-0.128$ & $-0.100\pm 0.0006$ & $1.398 \pm 0.001$& $<10$ $^{-4}$ \\
\hline
\end{tabular}
\end{table*}

Here we also present the relations between the \civ EW and the
continuum luminosity, the SMBH mass, and $\lb/\ledd$ in different
redshift bins, same to that in section 3.1 (see Table 3).

\begin{table*}
\centering \caption{Results of the samples in different redshift
bins for the relations between the \civ EW with the continuum
luminosity, the SMBH mass, and $\lb/\ledd$. Col.(1): different
relations; Col.(2): different redshift bins; Col.(3): number of
quasars in different redshift bins; Col.(4): correlation
coefficient of the linear fit; Col.(5): slope of the linear fit
together with its error; Col.(6): intercept of the linear fit with
its error at $\rm{log} L_ \lambda (1549\AA)=10^{42}$ \ergs,
$\rm{log} M_{\rm BH}/\Msun = 9$, $\rm{log} \lb/\ledd=-1$,
respectively; Col.(7): the possibility of null hypothesis. }
\begin{tabular}{lcccccccccc} \hline \hline
relations & z  & number  & R & slope & intercept & P \\
(1) & (2) & (3) & (4) & (5) & (6) & (7)\\
\hline
log \civ EW - log $ L_ \lambda$ (1549\AA)
& $1.5\le z \le 2.0$ & 7981 & $-0.484$ & $-0.244 \pm 0.0005$ & $ 1.936 \pm 0.032 $ & $<10$ $^{-4}$  \\
& $2.0 <  z \le 2.5$ & 3227 & $-0.452$ & $-0.234 \pm 0.0008$ & $ 1.987 \pm 0.048 $ & $<10$ $^{-4}$  \\
& $2.5 <  z \le 3.5$ & 1855 & $-0.513$ & $-0.251 \pm 0.0011$ & $ 2.103 \pm 0.066 $ & $<10$ $ ^{-4}$ \\
& $3.5 <  z \le 5.1$ &  897 & $-0.367$ & $-0.228 \pm 0.0025$ & $ 2.167 \pm 0.151 $ & $<10$ $^{-4}$  \\
\hline log \civ EW - log $ M_{\rm BH}/\Msun$
& $1.5\le z \le 2.0$ & 7891 & $-0.354$ & $-0.207 \pm 0.0006$ & $1.823  \pm 0.008  $ & $<10$ $^{-4}$ \\
& $2.0  < z \le 2.5$ & 3227 & $-0.412$ & $-0.234 \pm 0.0009$ & $1.855  \pm 0.011  $ & $<10$ $^{-4}$ \\
& $2.5  < z \le 3.5$ & 1855 & $-0.413$ & $-0.212 \pm 0.0011$ & $1.943 \pm 0.014 $ & $ <10$ $ ^{-4}$ \\
& $3.5  < z \le 5.1$ &  897 & $-0.294$ & $-0.142 \pm 0.0020$ & $2.015 \pm 0.025 $ & $ <10$ $^{-4}$\\
\hline log \civ EW - log $\lb/\ledd $
& $1.5\le z \le 2.0$ & 7981 & $-0.310$ & $-0.247 \pm 0.0009$ & $1.898 \pm 0.001 $ & $<10$ $^{-4}$ \\
& $2.0  < z \le 2.5$ & 3227 & $-0.148$ & $-0.123 \pm 0.0013$ & $1.890 \pm 0.002 $ & $<10$ $^{-4}$ \\
& $2.5  < z \le 3.5$ & 1855 & $-0.178$ & $-0.138 \pm 0.0017$ & $2.004 \pm 0.002 $ & $<10$ $^{-4}$ \\
& $3.5  < z \le 5.1$ &  897 & $ 0.016$ & $ 0.012 \pm 0.0030$ & $2.037 \pm 0.004  $ & $ 0.7182 $\\
\hline
\end{tabular}
\end{table*}

\section{conclutions}
Baldwin Effect has been explored for a long time and we are still
doing our efforts to explore it. In our paper, we discuss the
Baldwin Effect and its evolutionary effect from a large sample of
26623 quasars ($1.5 \le z\le 5.1$) from DR5. And also, using a
sample of 13960 quasars with broad \civ emission line from DR5, we
explore the physical parameters that drive the \civ Baldwin
Effect. The main conclusions can be summarized as follows: (1)
Baldwin Effect exists in the large sample of 26623 quasars from
DR5 of SDSS; (2) According to the slopes in our four
redshift-bins, up to $z\approx 5$, there's no evolutionary effect
for the slope of Baldwin Effect, however, these slopes are
relatively steeper than the slope for the total sample of 26623
quasars; (3) By the 13960 quasars with SMBH masses estimate, we
find that the relation between \civ EW and SMBH mass is stronger
than that between \civ EW and $\lb/\ledd $ (from $R=-0.344$ to
$R=-0.128$), SMBH mass seems to be a better underlying physical
parameter.

\section{ACKNOWLEDGMENTS}
We thank the anonymous referee for his/her comments and
instructive suggestions. This work has been supported by the
National Science Foundation of China (Nos. 10733010, 10778616,
10633020, 10403005, 10325313, 10233030 and 10521001), the
Science-Technology Key Foundation from Education Department of P.
R. China (No. 206053), and and the China Postdoctoral Science
Foundation (No. 20060400502).


\end{document}